\documentclass[twocolumn,prl,showpacs,preprintnumbers]{revtex4}
\usepackage[applemac]{inputenc}
\usepackage[T1]{fontenc} 
\usepackage{amsmath} 
\usepackage{amsfonts} 
\usepackage{amssymb}
\usepackage{graphicx} 
\usepackage[usenames]{color} 

\usepackage{bbm}
\def\beq{\begin{equation}}
\def\eeq{\end{equation}}
\def\bsp{\begin{split}}
\def\esp{\end{split}}
\def\bea{\begin{eqnarray}}
\def\eea{\end{eqnarray}}
\def\ba{\begin{array}}
\def\ea{\end{array}}

\def\lb{\left(}
\def\rb{\right)}

\def\l.{\left.}
\def\r.{\right.}

\def\part{\partial}

\begin{document}

\preprint{UdeM-GPP-TH-13-222}
\preprint{arXiv:1304.3734}
\title{Macroscopic quantum spin tunnelling with two interacting spins.}
\author{Solomon A. Owerre}
\email{solomon.akaraka.owerre@umontreal.ca}
\author{M. B. Paranjape} 
\email{paranj@lps.umontreal.ca}
\affiliation{Groupe de physique des particules, D\'epartement de physique,
Universit\'e de Montr\'eal,
C.P. 6128, succ. centre-ville, Montr\'eal, 
Qu\'ebec, Canada, H3C 3J7 }

\begin{abstract}

\section{Abstract}  
We study the simple Hamiltonian, $H=-K(S_{1z}^2 +S_{2z}^2)+ \lambda\vec S_1\cdot\vec S_2$, of two, large, coupled spins which are taken equal, each of total spin $s$ with $\lambda$ the exchange coupling constant.  The exact ground state of this simple Hamiltonian is not known for an antiferromagnetic coupling corresponding to the $\lambda>0$.  In the absence of the exchange interaction, the ground state is four fold degenerate, corresponding to the states where the individual spins are in their highest weight or lowest weight states, $|\hskip-1 mm\uparrow, \uparrow\rangle, |\hskip-1 mm\downarrow, \downarrow\rangle, |\hskip-1 mm\uparrow, \downarrow\rangle, |\hskip-1 mm\downarrow, \uparrow\rangle$, in obvious notation. The first two remain exact eigenstates of the full Hamiltonian.  However, we show the that   the two states $ |\hskip-1 mm\uparrow, \downarrow\rangle, |\hskip-1 mm\downarrow, \uparrow\rangle$  organize themselves into the combinations $|\pm\rangle=\frac{1}{\sqrt 2} (|\hskip-1 mm\uparrow, \downarrow\rangle \pm |\hskip-1 mm\downarrow \uparrow\rangle)$, up to perturbative corrections.    For the anti-ferromagnetic case, we show that the ground state is non-degenerate, and we find the interesting result that for integer spins the ground state is  $|+\rangle$,  and the first excited state is the anti-symmetric combination   $|-\rangle$ while for half odd integer spin, these roles are exactly reversed.  The energy splitting however, is proportional to $\lambda^{2s}$, as expected by perturbation theory to the $2s^{\rm th}$ order.  We obtain these results through the spin coherent state path integral.
\end{abstract}

\pacs{73.40.Gk,75.45.+j,75.50.Ee,75.50.Gg,75.50.Xx,75.75.Jn}

\maketitle

%%%%%%%%%%%%%%%%%%

{\it{Introduction}-}
We study the case of two large, coupled, quantum spins in the presence of a large, simple, easy axis anisotropy, interacting with each other through a  standard spin-spin exchange coupling, corresponding to the Hamiltonian
\beq
H=-K(S_{1z}^2 +S_{2z}^2)+ \lambda\vec S_1\cdot\vec S_2.\label{1}
\eeq 
We will consider $K>0$ and specialize to the case of equal spins $\vec S_1=\vec S_2=\vec S$.   $\lambda>0$ gives an anti-ferromagnetic coupling while  $\lambda<0$ sign corresponds to ferromagnetic coupling.  The spins $\vec S_i$ could correspond to quantum spins of macroscopic multi-atomic molecules \cite{A, B, F}, or the quantum spins a macroscopic ferromagnetic grains \cite{c}, or the average spin of each of the two staggered Ne\'el sub-lattices in a quantum anti-ferromagnet \cite{B,E}.  

The non-interacting system is defined by $\lambda=0$, here the spin eigenstates of $S_{iz}$, notationally $|s,s_{1z}\rangle\otimes|s,s_{2z}\rangle \equiv |s_{1z},s_{2z}\rangle$,  are obviously exact eigenstates.     The ground state is four-fold degenerate, corresponding to the states $ |s,s\rangle$, $ |-s,-s\rangle$, $ |s,-s\rangle$ and  $ |-s,s\rangle$, which we will write as   $|\hskip-1 mm\uparrow, \uparrow\rangle, |\hskip-1 mm\downarrow, \downarrow\rangle, |\hskip-1 mm\uparrow, \downarrow\rangle, |\hskip-1 mm\downarrow, \uparrow\rangle$, each with energy $E=-2Ks^2$. The first excited state, which is 8 fold degenerate, is split from the ground state by energy $\Delta E=K(2s-1)$.    

In the weak coupling limit, $\lambda/K\to 0$,  it is an interesting question to ask what is the ground state and the first few excited states of the system for large spin $\vec S$.  Surprisingly, this is yet, in general, an unsolved problem.    For spin $1/2$, the exact eigenstates are trivially found, for spin $1$, the problem is a $9\times 9$ matrix, which again can be diagonalized, but soon the problem becomes intractable.  In principle we must diagonalize a $(2s+1)^2\times (2s+1)^2$ matrix, that though is rather sparse, is not amenable to an exact diagonalization.    For weak coupling the  anisotropic  potential continues to align or anti-align the spins along the $z$ axis in the ground state.

As the non-interacting ground state is four fold degenerate, in first order degenerate  perturbation theory, we should diagonalize the exchange interaction in the degenerate subspace.  However, it turns out to be already diagonal in that subspace.  The full Hamiltonian can be alternatively written as
\beq
H=-K(S_{1z}^2 +S_{2z}^2)+ \lambda \left(S_{1z}S_{2z} +\frac{1}{2}(S_1^+S_2^-+S_1^-S_2^+)\right),\label{6}
\eeq
where $S_i^\pm=S_{ix}\pm iS_{iy}$ for $i=1,2$.  $S_i^\pm$ act as raising and lowering operators for $S_{iz}$, and hence they must annihilate the states $|\hskip-1 mm\uparrow, \uparrow\rangle, |\hskip-1 mm\downarrow, \downarrow\rangle$.  Thus the two states $|\hskip-1 mm\uparrow, \uparrow\rangle, |\hskip-1 mm\downarrow, \downarrow\rangle$  are actually exact eigenstates of the full Hamiltonian with exact energy eigenvalue  $(-2K+\lambda) s^2$.  These two states do not mix with  the two states $|\hskip-1 mm\uparrow, \downarrow\rangle, |\hskip-1 mm\downarrow, \uparrow\rangle$ as the eigenvalue of $S_{1z}+S_{2z}$, which is conserved, is respectively $+2s$, $-2s$ and 0.  The perturbation, apart from the diagonal term $\lambda S_{1z}S_{2z}$, acting on the two states $|\hskip-1 mm\uparrow, \downarrow\rangle, |\hskip-1 mm\downarrow, \uparrow\rangle$ takes them out of the degenerate subspace, thus this part does not give any correction to the energy.  The action of the diagonal term on either of these states is  equal to $-\lambda s^2$.  Thus the perturbation corresponds to the identity matrix within the degenerate subspace of the two states $|\hskip-1 mm\uparrow, \uparrow\rangle, |\hskip-1 mm\downarrow, \downarrow\rangle$, with eigenvalue $-\lambda s^2$.  This  yields, in first order degenerate perturbation theory, the perturbed energy eigenvalue of $(-2K-\lambda) s^2$ for the two states $|\hskip-1 mm\uparrow, \downarrow\rangle, |\hskip-1 mm\downarrow, \uparrow\rangle$.  Thus the following picture emerges of the first four levels in first order degenerate perturbation theory.  For the $\lambda<0$ (ferromagnetic coupling), the states  $|\hskip-1 mm\uparrow, \uparrow\rangle, |\hskip-1 mm\downarrow, \downarrow\rangle$ are the exact, degenerate ground states of the theory, with energy eigenvalue $(-2K+\lambda)s^2=(-2K-|\lambda|)s^2$.  The first excited states are also degenerate, but only within first order degenerate perturbation theory.  They are given by $|\hskip-1 mm\uparrow, \downarrow\rangle, |\hskip-1 mm\downarrow, \uparrow\rangle$, with energy eigenvalue $(-2K-\lambda)s^2=(-2K+|\lambda|)s^2$.  For the $\lambda>0$ (anti-ferromagnetic coupling), the roles are exactly reversed.  The states $|\hskip-1 mm\uparrow, \downarrow\rangle, |\hskip-1 mm\downarrow, \uparrow\rangle$ give the degenerate ground state with energy $(-2K-\lambda)s^2$
in first order degenerate perturbation, while the states $|\hskip-1 mm\uparrow, \uparrow\rangle, |\hskip-1 mm\downarrow, \downarrow\rangle$ give the exact, first (doubly degenerate) excited level with energy $(-2K+\lambda)s^2$.  

In this communication, we will show that in fact, the states $|\pm\rangle=\frac{1}{\sqrt 2} (|\hskip-1 mm\uparrow, \downarrow\rangle \pm |\hskip-1 mm\downarrow \uparrow\rangle)$ are the appropriate linear combinations  implied by the degenerate perturbation theory, for the ground state in the anti-ferromagnetic case, and they are the second and third excited states in the ferromagnetic case.  We will also show that the states $|\pm\rangle$ are no longer degenerate.  The perturbing Hamiltonian links the state $|\pm s, \mp s\rangle$ only to the state $|\pm s\mp 1,\mp s\pm 1\rangle$.  To reach the state $|\mp s,\pm s\rangle$ from the state $|\pm s,\mp s\rangle$ requires one to go to $2s^{\rm  th}$ order in perturbation, and $s$ is assumed to be large.  Indeed, we find our results via macroscopic quantum tunnelling using the spin coherent state path integral.  Using the path integral to determine large orders in perturbation theory has already been studied in field theory \cite{cl}.

 {\it Spin coherent state path integral -} The quantum (large) spin systems can be described by the spin coherent state path integral \cite{l, D, AB}. 
 \beq
\langle \chi |e^{-\beta H}|\psi\rangle ={\cal N}\int_{\psi}^{\chi}\hskip-2mm {\cal D}\theta_i{\cal D}\phi_i \,\, e^{-S_E}.\label{fi}
\eeq
$S_E$ is the Euclidean action which corresponding to dynamics of  particles moving on a two sphere, and which contains  first order kinetic term, the Wess-Zumino-Novikov-Witten (WZNW) term for the spin degree of freedom \cite{wznw}.   The WZNW term for a spin degree of freedom can be written in a parametrisation independent fashion by extending the time dimension by an additional spatial dimension denoted by $x$.     Then the WZNW term corresponds to the integral
\beq
S_{WZNW}=\sigma\int dt\int_0^1dx \hat S(t,x)\cdot(\partial_x \hat S(t,x)\times \partial_t \hat S(t,x)).\label{wz}
\eeq
where $\hat S(t,x)$ is a  3-vector of unit norm, which satisfies at $x=0$ the boundary condition $\hat S(t,0)=\hat S(t)$, and at $x=1$  that the spin configuration is constant, which we can take $\hat S(t,1)=\hat z$.  It does not actually matter how the spin configuration is extended into the extra dimension, as long as the boundary conditions are respected, the integral Eq.\eqref{wz} changes only by an integer multiple of $4\pi$.  Thus taking $\sigma=N/2$ where $N$ is an integer, means that this discrete ambiguity is unobservable in the functional integral \eqref{fi}, and nicely gives us the quantization of the spin.  We refer the reader to \cite{wznw} for all the details.  If we  parametrize the configuration explicitly as $\hat S(t,x)= (\sin((1-x)\theta(t))\cos\phi(t), \sin((1-x)\theta(t))\sin\phi(t),\cos((1-x)\theta(t)))$ which satisfies the boundary conditions at $x=0$ and $x=1$, then after an easy calculation of the integrand we find that the $x$ integration can be explicitly done giving
\bea
S_{WZNW}&=&\int dt\int_0^1dx -\sigma\dot\phi(t)\sin((1-x)\theta(t))\nonumber\\
&=&\int dt\left. -\sigma\dot\phi(t)\cos((1-x)\theta(t))\right|_0^1\nonumber\\
&=&\int dt -\sigma\dot\phi(t)(1-\cos(\theta(t))).
\eea
Hence we recover the familiar expression in condensed matter physics for the the Wess-Zumino-Novikov-Witten term.

Our two spin system, in real time, is governed by an action  $S= \int dt\mathcal{L}$ where, 
\beq 
\begin{split}
 \mathcal{L}=\int  dx \thinspace \sigma_1\hat S_1\cdot (\partial_x\hat S_1 \times  \partial_t\hat S_1)- V_1(\hat S_1)+\\ +\int  dx \thinspace \sigma_2\hat S_2 \cdot (\partial_x{\hat S_2} \times\partial_t\hat S_2)- V_2(\hat S_2)- \lambda\hat S_1\cdot \hat S_2
\label{2}
\end{split}
\eeq
where now $\hat S_i= \left(\sin \theta_i \cos \phi_i, \sin \theta_i \sin \phi_i, \cos \theta_i\right), i =1,2$ are two different 3-vectors of unit norm, representing semi-classically the quantum spin \cite{c} and $\sigma_i$ are the values of each spin.  In terms of spherical coordinates the Lagrangian takes the form
\bea
 \mathcal{L}&=&-\sigma_1 \dot{\phi_1}(1-\cos\theta_1) -V_1(\theta_1,\phi_1)\nonumber\\&-&\sigma_2 \dot{\phi_2}(1-\cos\theta_2)-V_2(\theta_2,\phi_2)\nonumber\\&-& \lambda \left(\sin \theta_1\sin \theta_2 \cos(\phi_1-\phi_2)+\cos \theta_1\cos \theta_2\right).
\label{3}
\eea

We consider the special case of equal spins, with $\sigma_1 =\sigma_2 =s$.  Our analysis is valid if we restrict our attention to any external potential with easy-axis, azimuthal symmetry, with  a reflection symmetry (along the azimuthal axis), as in \cite{am}, $V_i(\theta_i,\phi_i)\equiv V(\theta_i)=V(\pi-\theta_i),    i =1,2$.   The potential is further assumed to have a minimum at the north pole and the south pole, at $\theta_i=0$, and $\pi$.  In our case the potential  is explicitly
\beq
V(\hat S_i)\equiv V(\theta_i,\phi_i)= K  \sin^2\theta_i\label{potential}.
\eeq 
It was shown in Ref. \cite{am}, for uncoupled spins, that  quantum tunnelling between the spin up and down states of each spin separately is actually absent  because of conservation of the $z$ component of each spin.  With the exchange interaction only the total $z$ component is conserved allowing transitions $|\hskip-1mm\uparrow,\downarrow\rangle\longleftrightarrow |\hskip-1mm\downarrow,\uparrow\rangle$.      In general tunnelling exists if there is an equipotential path that links the beginning and end points.   We will see that such an equipotential path exists, but through complex values of the phase space variables.  

We must find the critical points of the  Euclidean action with $t\rightarrow -i\tau$, see Ref. \cite{sc}, which gives
\bea
& \mathcal{L}_E=is \dot{\phi_1}(1-\cos\theta_1) +V(\theta_1)+is \dot{\phi_2}(1-\cos\theta_2)+V(\theta_2)\nonumber\\&+ \lambda \left(\sin \theta_1\sin \theta_2 \cos(\phi_1-\phi_2)+\cos \theta_1\cos \theta_2\right).
\label{5}
\eea
The solutions must start at $(\theta_1,\phi_1)=(0,0)$ and $(\theta_2, \phi_2)=(\pi,0)$, say, and evolve to $(\theta_1, \phi_1)=(\pi, 0)$ and $(\theta_2, \phi_2)=(0,0)$.  In Euclidean time, the WZNW term has become imaginary and the equations of motion in general only have solutions for complexified field configurations.  Varying with respect to $\phi_i$ gives equations that correspond to the conservation of angular momentum:
\bea
is  \frac{d}{d\tau}\lb 1-\cos\theta_1\rb +\lambda \sin \theta_1 \sin \theta_2 \sin\lb \phi_1 -\phi_2 \rb =0
\label{7}
\eea
\bea
is  \frac{d}{d\tau}\lb 1-\cos\theta_2 \rb -\lambda \sin \theta_1 \sin \theta_2 \sin\lb \phi_1 -\phi_2 \rb =0
\label{8}
\eea
Varying with respect to $\theta_i$ gives the equations: 
\bea
 \frac{\partial \mathcal{L}_E}{\partial \theta_1} =0 =\frac{\partial \mathcal{L}_E}{\partial \theta_2}
\label{9}
\eea
Adding  Eqn's \eqref{7} and  \eqref{8}  we simply get
\bea
\frac{d}{d\tau}\lb \cos\theta_1 +\cos\theta_2\rb  =0.
\label{10}
\eea
Hence $ \cos\theta_1 +\cos\theta_2 = l=0 \quad\Longrightarrow \theta_2 = \pi -\theta_1$, where the constant $l$ is chosen to be zero using the initial condition $ \theta_1 =0,  \theta_2 = \pi$.   We can now eliminate $\theta_2$ from the equations of motion and writing $\theta = \theta_1 $, $\phi=\phi_1-\phi_2$ and  $\Phi=\phi_1+\phi_2$ and taking $V_i(\theta_i)=V(\theta_i)=V(\pi-\theta_i)$ we get  the effective Lagrangian: 
\beq 
\begin{split}
 \mathcal{L}=is \dot{\Phi} -is \dot{\phi}\cos\theta+U(\theta, \phi)
\label{12}
\end{split}
\eeq
where $ U\lb\theta, \phi\rb= 2V\lb\theta\rb+ \lambda \lb\sin^2\theta\cos\phi - \cos^2\theta \rb+\lambda$ is the effective potential energy. We have added a constant $\lambda$ so that the potential is normalized to zero at $\theta =0$. The first term in the Lagrangian is a total derivative and drops out. The equations of motion become:
\bea
  is\dot{\phi} \sin\theta &=& -\frac{\partial U\lb\theta, \phi\rb}{\partial \theta}
\label{14}\\
is\dot{\theta} \sin\theta &=& \frac{\partial U\lb\theta, \phi\rb}{\partial \phi}
\label{15}
\eea
These equations have no solutions on the space of real functions $\theta(\tau), \phi(\tau)$ due to the explicit $i$ on the left hand side.  The analog of conservation of energy follows immediately from these equations, this is easily derived by multiplying \eqref{14} by $\dot{\theta}$ and \eqref{15} by $\dot{\phi}$  and subtracting,  giving:
\beq 
\frac{d U\lb\theta, \phi\rb}{d\tau}= 0 \quad  \text{i.e},\quad
U\lb\theta, \phi\rb = const. = 0
\label{16}
\eeq
The constant has been set to $0$ using the initial condition $\theta = 0$.  Thus we have, specializing to our case Eqn. \eqref{potential}
\beq
 U\lb\theta, \phi\rb= (2 K + \lambda \lb\cos\phi+1\rb)\sin^2\theta =0
\eeq
implying $(2 K + \lambda \lb\cos\phi+1\rb)=0$ since $\sin^2\theta\ne 0$,  is required for a non-trivial solution.  Thus
 \beq
\cos\phi=-\lb\frac{2K}{\lambda}+1\rb      
\label{20}
\eeq
and we see that $\phi$ must be a constant.  This is not valid in general, it is due to the specific choice of the external potential Eqn. \eqref{potential}.  Since $K>|\lambda|$ we get $|\cos\phi|>1$, which of course has no solution for real $\phi$.  We take $\phi=\phi_R+i\phi_I$ which gives $\cos\phi=\cos\phi_R\cosh\phi_I-i\sin\phi_R\sinh\phi_I$. As the RHS of Eqn. \eqref{20} is real, we must have either $\phi_I=0$ or $\phi_R=n\pi$ or both.  Clearly the $\phi_I=0$ cannot yield a solution for Eqn. \eqref{20}, hence we must have $\phi_R=n\pi$.  As we must impose $2\pi$ periodicity on $\phi_R$ only $n=0$ or 1 exist. Then we get
 \begin{align}
 \cos\phi=(-1)^n\cosh\phi_I&=
  \begin{cases}
   -\lb\frac{2K}{\lambda}+1\rb        & \text{if } \lambda > 0 \\
    +\lb\frac{2K}{|\lambda|}-1\rb        & \text{if } \lambda < 0\\
  \end{cases}\label{21}
 \end{align}
 Thus $n=1$ for $\lambda>0$ and $n=0$ for $\lambda<0$ allowing for the unified expression
 \beq
 \cosh\phi_I=\frac{2K+\lambda}{|\lambda|}.\label{23}
 \eeq
Eqn. \eqref{15}  simplifies to
\beq
is\frac{\dot\theta}{\sin\theta}=-\lambda\sin\phi=-i\lambda (-1)^n\sinh\phi_I=i|\lambda|\sinh\phi_I\label{14a}
\eeq
as $\lambda (-1)^n=-|\lambda|$.  Eqn. \eqref{23} has two solutions  positive $\phi_I$ corresponds to the instanton, ($\dot\theta>0$), and negative $\phi_I$  corresponds to the anti-instanton, ($\dot\theta<0$).  
The equation is trivially integrated with solution
\beq 
\theta\lb \tau\rb =  2 \arctan \lb e^{\omega (\tau-\tau_0)}\rb
\label{18}
\eeq
 where $\omega= (|\lambda|/s)\sinh\phi_I$ and at $\tau=\tau_0$ we have $\theta(\tau)= \pi/2$.  Thus $\theta(\tau)$ interpolates  from 0 to $\pi$ as $\tau=-\infty\rightarrow\infty$ for the instanton and from $\pi$ to 0 for an anti-instanton.  

Using $\dot\phi=0$ and \eqref{16} we see that the action for this instanton trajectory, let us call it $S_0$, simply vanishes $S_0=\int_{-\infty}^{\infty} d\tau \mathcal{L}= 0.$
So where could the amplitude come from?  We have not taken into account the fact  that $\phi$ must be translated from $\phi=0$ (any initial point will do, as long as it is consistently used to compute the full amplitude) to $\phi=n\pi+i\phi_I$ before the instanton can occur and then back to $\phi=0$ after the instanton has occurred.  Normally such a translation has no effect, either the change at the beginning cancels that at the end, or if the action is second order in time derivative,  moving adiabatically gives no contribution.  But in the present case, before the instanton occurs, $\theta=0$, but after it has occurred, $\theta=\pi$.  As $\dot\phi$ is multiplied by $\cos\theta$ in the action, the two contributions actually add,  there is a net contribution to the action.  Indeed the change of the full action for the combination of the instanton and the changes in $\phi$ is given by
\bea
\Delta S&=&\int_0^{n\pi+i\phi_I} \hskip-.8cm -is d\phi \cos\theta|_{\theta=0} +S_0+\int^0_{n\pi+i\phi_I}  \hskip-.8cm -is d\phi \cos\theta|_{\theta=\pi}\nonumber \\
&=&-is2n\pi +2s\phi_I.
\eea

We will use this information to compute the following matrix element, using the spin coherent states $|\theta,\phi\rangle$ and the lowest two energy eigenstates $|E_0\rangle$ and $|E_1\rangle$:
\bea
\langle \theta_f,\phi_f|e^{-\beta H}|\theta_i,\phi_i\rangle =e^{-\beta E_0}\langle \theta_f,\phi_f|E_0\rangle\langle E_0|\theta_i,\phi_i\rangle\nonumber\\ 
+e^{-\beta E_1}\langle \theta_f,\phi_f|E_1\rangle\langle E_1|\theta_i,\phi_i\rangle +\cdots
\eea
On the other hand, the matrix element is given by the spin coherent state path integral
\beq
\langle \theta_f,\phi_f|e^{-\beta H}|\theta_i,\phi_i\rangle ={\cal N}\int_{\theta_i,\phi_i}^{\theta_f,\phi_f}\hskip-2mm {\cal D}\theta{\cal D}\phi \,\, e^{-S_E}.
\eeq
The integration is done in the saddle point approximation.  With $(\theta_i,\phi_i)=(0,0)$ corresponding to the state $|\hskip-1mm\uparrow,\downarrow \rangle $ and  $(\theta_f,\phi_f)=(\pi,0)$ corresponding to the state $|\downarrow,\uparrow\rangle$we get, with a mild abuse of notation
\beq
\langle\downarrow,\uparrow| e^{-\beta H}|\hskip-1mm\uparrow,\downarrow \rangle ={\cal N}e^{-\Delta S}\kappa \beta(1+\cdots)
\eeq
where $\kappa$ is the ratio of the square root of the determinant of the operator governing the second order fluctuations  about the instanton excluding the time translation zero mode, and that of the free determinant.  It can in principle be calculated, but we have not done this.  The zero mode is taken into account by integrating over the position of the occurrence of the instanton giving rise to  the  factor of $\beta$.   $\cal N$ is the overall normalisation including the square root of the free determinant which is given by $Ne^{-E_0\beta}$ where $E_0$ is the unperturbed ground state energy and $N$ is a constant from the ground state wave function.  The result exponentiates, but since we must sum over all sequences of one instanton followed by any number of anti-instanton/instanton pairs, the total number of instantons and anti-instantons is odd, and we get
\beq
e^{-\Delta S}\kappa\beta\to\sinh\lb e^{-\Delta S}\kappa\beta\rb
\eeq
Given $\Delta S=-is2n\pi +2s\phi_I$ and solving Eqn.  \eqref{23} for $\phi_I$ for $K\gg|\lambda|$ 
\beq
\phi_I={\rm arccosh}\lb\frac{2K+\lambda}{|\lambda|}\rb\approx\ln\lb\frac{4K}{|\lambda|}\rb
\eeq
gives 
\footnote{We thank Sung-Sik Lee for a discussion at this point which led to the understanding of the (expected to him) following result.}:
 \begin{align}
e^{-\Delta S}&=
  \begin{cases}
  e^{is2\pi -2s\phi_I}      & \text{if } \lambda > 0=\begin{cases}\lb\frac{|\lambda|}{4K}\rb^{2s}&\hskip-5mm\text{if }s\in\mathbf{Z}\\ -\lb\frac{|\lambda|}{4K}\rb^{2s}&\hskip-5mm\text{if }s\in\mathbf{Z}+1/2\end{cases} \\
 \lb\frac{|\lambda|}{4K}\rb^{2s}     & \text{if } \lambda < 0\\
  \end{cases}\label{21a}
 \end{align}
Then we get
\beq
\langle\downarrow,\uparrow| e^{-\beta H}|\hskip-1mm\uparrow,\downarrow \rangle =\pm\lb\frac{1}{2}e^{\lb\frac{|\lambda|}{4 K }\rb^{2s}\kappa\beta}-\frac{1}{2}e^{-\lb\frac{|\lambda|}{4 K }\rb^{2s}\kappa\beta}\rb Ne^{-\beta E_0}
\eeq
where the $-$ sign only applies for the case of anti-ferromagnetic coupling with half odd integer spin, $\lambda>0, s={\mathbf Z}+1/2$.  
An essentially identical analysis yields
\bea
&&\hskip-1cm\langle\downarrow,\uparrow| e^{-\beta H}|\hskip-1mm\downarrow,\uparrow \rangle =\langle\uparrow,\downarrow| e^{-\beta H}|\hskip-1mm\uparrow,\downarrow \rangle\nonumber\\
&=&\lb\frac{1}{2}e^{\lb\frac{|\lambda|}{4 K }\rb^{2s}\kappa\beta}+\frac{1}{2}e^{-\lb\frac{|\lambda|}{4 K }\rb^{2s}\kappa\beta}\rb Ne^{-\beta E_0}.
\eea

These calculated matrix elements should now be compared with what is expected for the exact theory:
\bea
\langle\downarrow,\uparrow| e^{-\beta H}|\hskip-1mm\uparrow,\downarrow \rangle&=&e^{-\beta(E_0-\frac{1}{2}\Delta E)}\langle\downarrow,\uparrow|  E_0\rangle\langle E_0 |\hskip-1mm\uparrow,\downarrow \rangle\nonumber \\
&+&e^{-\beta(E_0+\frac{1}{2}\Delta E)}\langle\downarrow,\uparrow|  E_1\rangle\langle E_1|\hskip-1mm\uparrow,\downarrow \rangle\nonumber\\
\eea
and say
\bea
\langle\downarrow,\uparrow| e^{-\beta H}|\hskip-1mm\downarrow,\uparrow \rangle&=&e^{-\beta(E_0-\frac{1}{2}\Delta E)}\langle\downarrow,\uparrow|  E_0\rangle\langle E_0 |\hskip-1mm\downarrow,\uparrow \rangle\nonumber \\
&+&e^{-\beta(E_0+\frac{1}{2}\Delta E)}\langle\downarrow,\uparrow|  E_1\rangle\langle E_1|\hskip-1mm\downarrow,\uparrow \rangle\nonumber\\
\eea

The energy splitting can be read off from this result
\beq
\Delta E=E_1-E_2=2\lb\frac{|\lambda|}{4 K }\rb^{2s}\kappa
\eeq
and our main result follows, the low energy eigenstates are given by
\beq
| E_0\rangle =\frac{1}{\sqrt 2}(|\hskip-1mm\downarrow,\uparrow \rangle +|\hskip-1mm\uparrow,\downarrow \rangle)\quad\quad | E_1\rangle =\frac{1}{\sqrt 2}(|\hskip-1mm\downarrow,\uparrow \rangle -|\hskip-1mm\uparrow,\downarrow \rangle)
\eeq
for $\lambda<0$ (although here the energy eigenstates should be $| E_3\rangle$ and $|E_4\rangle$) and $\lambda>0$ for $s\in\mathbf Z$, while for the anti-ferromagnetic $\lambda>0$ case with $s\in\mathbf{Z}+1/2$ we get

\beq
| E_0\rangle =\frac{1}{\sqrt 2}(|\hskip-1mm\downarrow,\uparrow \rangle -|\hskip-1mm\uparrow,\downarrow \rangle)\quad\quad | E_1\rangle =\frac{1}{\sqrt 2}(|\hskip-1mm\downarrow,\uparrow \rangle +|\hskip-1mm\uparrow,\downarrow \rangle).
\eeq
 This understanding of the ground state in the anti-ferromagnetic case is our main result. This difference in the ground states for integer and half odd integer spins is understood in terms of the Berry phase \cite{l} (computed by the change in the Wess-Zumino term) for the evolution corresponding to the instanton.  It can also be understood by looking  at perturbation theory to order $2s$, the details cannot be given here.  Briefly, one finds that  the effective $2\times 2$ Hamiltonian for the degenerate subspace is proportional to the identity plus off diagonal terms that are symmetric.  For the integer spin case the off diagonal terms are negative and for the half odd integer case they are positive.  Diagonalizing this $2\times 2$ matrix gives the solutions for the ground states, exactly as we have found.

{\it Conclusions-}
We have found the low energy eigenvalues and the corresponding eigenstates for the Hamiltonian of two equal, large, spins interacting with an easy axis anisotropy and a standard exchange interaction, the latter which is considered as a perturbation.  We find that the two states $|\hskip-1mm\downarrow,\uparrow \rangle ,|\hskip-1mm\uparrow,\downarrow \rangle$ reorganize into the symmetric and the anti-symmetric superposition because of quantum tunnelling transitions.  These transitions correspond to the $2s^{\rm th}$ order effects in perturbation theory.  The symmetric combination is the lower energy state for integer spin while the anti-symmetric state is the the lower energy state for half odd integer spins.  These states are respectively the ground states for an anti-ferromagnetic coupling.

{\it  Acknowledgments-} We thank Ian Affleck, Sung-Sik Lee and Cliff Burgess for useful discussions and  we thank  NSERC of Canada for financial support. 
%\vfill

%%%%%%%%%%%%%%%%%%%%  BIBLIO  %%%%%%%%%%%%%%%%%%%%%%%%%%%

\end{document}